# Technology Use in the Virtual R&D Teams

**Nader Ale Ebrahim*, Shamsuddin Ahmed**
Department of Engineering Design and Manufacture,
Faculty of Engineering, University of Malaya
Kuala Lumpur, Malaysia

**Salwa Hanim Abdul Rashid**

Centre for Product Design and Manufacturing
Department of Engineering Design and Manufacture
Faculty of Engineering, University of Malaya
Kuala Lumpur, Malaysia

**Zahari Taha**

Faculty of Manufacturing Engineering and Management Technology,
University Malaysia Pahang,
26300 Gambang,
Pahang, Malaysia

**Abstract:**

**Problem statement:** Although, literature proves the importance of the technology role in the effectiveness of virtual research and development (R&D) teams for new product development. However, the factors that make technology construct in a virtual R&D team are still ambiguous. The manager of virtual R&D teams for new product development does not know which type of technology should be used.

**Approach:** To address the gap and answer the question, the study presents a set of factors that make a technology construct. The proposed construct modified by finding of the field survey (N = 240). We empirically examine the relationship between construct and its factors by employing the Structural Equation Modeling (SEM). A measurement model built base on the 19 preliminary factors that extracted from literature review. The result shows 10 factors out of 19 factors maintaining to make technology construct.

**Result:** These 10 technology factors can be grouped into two constructs namely Web base communication and Web base data sharing. The findings can help new product development managers of enterprises to concentrate in the main factors for leading an effective virtual R&D team. In addition, it provides a guideline for software developers as well.

**Conclusion:** The second and third generation technologies are now more suitable for developing new products through virtual R&D teams.

**Key words:** Collaboration teams, questionnaires performance, cross-functional teams, product development, structural equation modeling, measurement model, literature review,e virtual,

## 1   INTRODUCTION

A virtual team is defined as "a small temporary group of geographically, organizationally and/or time dispersed knowledge workers who coordinate their work, mainly with electronic information and communication technologies to carry out one or more organization tasks" (Ale Ebrahim et al., 2009b). Virtual R&D team is a form of a virtual team, which includes the features of virtual teams and concentrates on R&D activities (Ale Ebrahim et al., 2011).

The members of a virtual R&D team use different types of communication technology to complete the research beyond space, time and organizational boundaries (Ale Ebrahim et al., 2010). "We are becoming more virtual all the time!" is heard in many global corporations today (Chudoba et al., 2005). On the other hand, new product development (NPD) is widely recognized as a key to corporate prosperity (Lam et al., 2007). The specialized skills and talents needed for developing new products often remain locally in pockets of excellence around the company. Therefore, enterprises, have no choice but to disperse their new product development units to gain access into such dispersed knowledge and skills (Kratzer et al., 2005). As a result, enterprises are finding that internal development of all technologies needed for new products and processes are difficult or impossible. They must increasingly receive technology from external sources (Stock and Tatikonda, 2004).

\* : Corresponding Author e-mail: aleebrahim@siswa.um.edu.my

Virtualization in NPD has recently begun to make a serious headway due to the rapid growth of a large variety of technologies. This means that virtuality in NPD is now technically possible (Leenders et al., 2003). Due to increasing and changing product features, by-and-large product development has become more complex, with increasing complexity in the supply chain. Therefore, more close collaboration between customers, developers, and suppliers has become vital. The foretold collaborations often involve individuals from different geographical locations that could now be brought together by using the various types of information technology (IT). IT offers a large number of benefits (Anderson et al., 2007). Although the use of the Internet for many purposes has received notable attention in the literature, little has been said about collaborative tool and effective virtual teams for NPD (Ale Ebrahim et al., 2009a). In addition, the literature did not reveal adequate focus on the factors which can construct a technological niche for a virtual R&D team for NPD. This aims to such a technological construct.

This paper is structured as follows. Firstly, based on prior research, we extracted the 19 factors of technology construct in the virtual R&D teams. Next, Structural Equation Modeling (SEM) was used as an analytical tool for testing the estimations and testing the technology construct measurement models. Then, we adjusted the preliminary technology construct model by fitting the model according to the SEM fitness indices and made a final measurement model. The paper infers with a discussion and future guidelines.

## 2       LITERATURE REVIEW

Virtual teams use digital communications, video and audio links, electronic whiteboards, e-mails, instant messaging, websites, chat rooms, etc. as substitutes for physical collocation of the team members (Baskerville and Nandhakumar, 2007, Pauleen and Yoong, 2001). Simple transmission of information from location *A* to another location *B* is not enough. However, a virtual environment presents significant challenges to effective communication (Walvoord et al., 2008). Being equipped with even the most advanced technologies is not necessarily sufficient to make a virtual team effective, since the internal group dynamics and external support mechanisms must also be present for a team to succeed in the virtual world (Lurey and Raisinghani, 2001). Virtual teams are technology-mediated groups of people from different disciplines that work on common tasks (Dekker et al., 2008) and therefore, the way the information technology is implemented seems to make the virtual teams outcome more or less likely (Anderson et al., 2007). The virtual R&D team's instructor should choose the appropriate technology based on the purpose of the team (Ale Ebrahim et al., 2009d).

The factors which make technology construct in a virtual R&D team are still not clearly set in the literature. We extracted 19 important factors related to the technology construct, based on a comprehensive review on technology view in the virtual R&D team working. Table 1summarizes the factors and their supported references. E-mails and conference calls are generally known as first generation technologies whereas online discussion boards, Power Point presentations, video tools and online meeting tools are second-generation technologies. Third generation technology refers typically to web-enabled shared workspaces with the Intranet or Internet (Lee-Kelley and Sankey, 2008).

Table 1 Summary of the factors related to technology construct in virtual teams

| Factor name | Factor descriptions | References |
| --- | --- | --- |
| **Tech1** | Use internet and electronic mail | (Redoli et al., 2008, Pauleen and Yoong, 2001, Lee-Kelley and Sankey, 2008, Thissen et al., 2007, Townsend et al., 1998) |
| **Tech2** | Online meeting on need basis | (Chen et al., 2007, Lee-Kelley and Sankey, 2008, Pena-Mora et al., 2000, Thissen et al., 2007) |
| **Tech3** | Web conferencing | (Coleman and Levine, 2008, Thissen et al., 2007, Zemliansky and Amant, 2008, Ale Ebrahim et al., 2009d) |
| **Tech4** | Seminar on the Web | (Zemliansky and Amant, 2008) |
| **Tech5** | Shared work spaces | (Lee-Kelley and Sankey, 2008) |
| **Tech6** | Video conferencing | (Chen et al., 2007, Zemliansky and Amant, 2008, Townsend et al., 1998) |
| **Tech7** | Audio conferencing | (Chen et al., 2007, Lee-Kelley and Sankey, 2008, Zemliansky and Amant, 2008) |
| **Tech8** | Online presentations | (Lee-Kelley and Sankey, 2008, Townsend et al., 1998) |
| **Tech9** | Share documents (off-line) | (Coleman and Levine, 2008, Ale Ebrahim et al., 2009d, Townsend et al., 1998) |
| **Tech10** | Share what is on your computer desktop with people in other locations (Remote access and control) | (Thissen et al., 2007, Ale Ebrahim et al., 2009c, Townsend et al., 1998) |
| **Tech11** | Do not install engineering software (get service through web browser) | (Coleman and Levine, 2008, Kotelnikov, 2007, Shumarova, 2009) |
| **Tech12** | Access service from any computer (in Network) | (Thissen et al., 2007, Shumarova, 2009) |
| **Tech13** | Standard phone service and hybrid services | (Thissen et al., 2007, Ale Ebrahim et al., 2009d, Townsend et al., 1998) |
| **Tech14** | Access shared files anytime, from any computer | (Lee-Kelley and Sankey, 2008, Townsend et al., 1998) |
| **Tech15** | Web database | (Coleman and Levine, 2008, Zemliansky and Amant, 2008, Ale Ebrahim et al., 2009d, Townsend et al., 1998) |
| **Tech16** | Provide instant collaboration | (Coleman and Levine, 2008, Thissen et al., 2007) |
| **Tech17** | Software as a service (canceling the need to install and run the application on the own computer) | (Coleman and Levine, 2008, Thissen et al., 2007, Townsend et al., 1998) |
| **Tech18** | Virtual research center for product development | (Zemliansky and Amant, 2008, Townsend et al., 1998) |
| **Tech19** | Integratable/compatible with the other tools and systems | (Coleman and Levine, 2008, Kotelnikov, 2007, Townsend et al., 1998) |

## 3 RESEARCH METHODOLOGY AND DATA COLLECTION

To build a measurement model of information technology construct in virtual R&D teams for new product development, we conducted a Web-based survey mainly in Malaysian and Iranian manufacturing enterprises, in a random sample of small and medium enterprises. Web-based survey method was selected because it is a cost-effective and quick method to obtain feedbacks from the beliefs of the respondents. The rapid expansion of Internet users has given Web-based surveys the potential to become a powerful tool in survey research (Sills and Song, 2002, Ebrahim et al., 2010). A Likert scale from one to five was used. This set-up provided the respondents with a series of attitude dimensions. For each factor, the respondents were asked whether the factor is unimportant or extremely important using a Likert scale rating. The questionnaires were e-mailed to the managing director, R&D manager, new product development manager, project and design manager and appropriate personnel who were most familiar with the R&D activities within the firm.

Invitation e-mails were sent to each respondent, reaching 972 valid email accounts, with reminders following every two weeks up to three months. 240 enterprises completed the questionnaire, for an overall response rate of 24.7% (Table 2).

Table 2 Summary of online survey data collection

| | |
|---|---|
| Numbers of e-mails sent to enterprises | 3625 |
| Total responses (Clicked the online web page) | 972 |
| Total responses / received questionnaire (%) | 26.8 |
| Total completed | 240 |
| Total completed / received questionnaire (%) | 24.7 |

## 4 ANALYSIS AND RESULTS

Gerbing and Anderson (1988) suggested using confirmatory factor analysis (CFA) for scale development because it affords stricter interpretation of uni-dimensionality than what is provided by traditional approaches such as coefficient alpha, item-total correlations, and exploratory factor analysis. The evidence that the measures were uni-dimensional, whereby a set of indicators (factors) shares only a single underlying construct, was assessed using CFA (Anderson and Gerbing, 1988). According to Anderson and Gerbing (1988), after data collection, the measures' purification procedures should be used to assess their reliability, uni-dimensionality, discriminant validity, and convergent validity. For reliability analysis, Cronbach's alpha was employed to each factor. From Table 3, all items with Cronbach's α greater than the threshold value of 0.6 were included in the analysis and the rest were omitted from analysis. Hence, the factors Tech1, Tech10, Tech11 and Tech13 were excluded from further analysis. In general, the reliability of the contents in the questionnaire exhibits good reliability across the samples.

Structural Equation Modeling (SEM) using AMOS 18 was employed for validating the measurement model. The statistical analysis were estimated simultaneously for both measurement and structural models (Dibrell et al., 2008). In order to ensure that the factors made the right construct, the measurement model was examined for its fit. Given this, the model was assessed for convergent and discriminant validity.

Convergent validity was established using a calculation of the factor loading, average variance extracted (AVE) and composite reliability (CR). The factors which have standardized loadings exceeding 0.50, were retained (Dibrell et al., 2008). The initial measurement model consisted of 19 factors (Tech1 to Tech19). After revising the measurement model by deleting Tech1, Tech10, Tech11 and Tech13, the AVE and CR were calculated. AVE larger than 0.5 is the threshold (McNamara et al., 2008). CR was calculated by squaring the sum of loadings, followed by division with the sum of squared loadings, plus the sum of the measurement error (Lin et al., 2008). CR should be greater than 0.6 (Huang, 2009). The measurement model had acceptable convergent validity since the calculated CR and AVE were 0.930 and 0.613, respectively.

Table 3 Summary of the final measures and reliabilities

| Factor name | Corrected Item-Total Correlation | Cronbach's Alpha if Item Deleted |
|---|---|---|
| Tech1 | 0.525 | 0.943 |
| Tech2 | 0.755 | 0.939 |
| Tech3 | 0.777 | 0.939 |
| Tech4 | 0.717 | 0.940 |
| Tech5 | 0.759 | 0.939 |
| Tech6 | 0.722 | 0.940 |
| Tech7 | 0.731 | 0.939 |
| Tech8 | 0.780 | 0.939 |
| Tech9 | 0.610 | 0.942 |
| Tech10 | 0.576 | 0.942 |
| Tech11 | 0.571 | 0.943 |
| Tech12 | 0.686 | 0.940 |
| Tech13 | 0.519 | 0.943 |
| Tech14 | 0.624 | 0.941 |
| Tech15 | 0.696 | 0.940 |
| Tech16 | 0.642 | 0.941 |
| Tech17 | 0.678 | 0.940 |
| Tech18 | 0.649 | 0.941 |
| Tech19 | 0.615 | 0.942 |

For discriminant validity, we used AMOS software using the Maximum Likelihood method (ML). The fitting indices were checked with their respective acceptance values (Table 4). We ran the AMOS for the model Ver1 (information technology construct with 15 factors), and found a non-significant chi-square value per degree of freedom (CMIN/DF = 7.232). Most of the remaining fit indices were not within the acceptable range. Thus, referring to the AMOS modification indices (MI), some of the factors which had the lowest factor loading or the same effect of remaining factor, were deleted. With this modification, the measurement model Ver2 had a significant chi-square per degrees of freedom (CMIN/DF = 4.767); other fit indices, RMSEA, RMR, and GFI were also in the acceptable range. Therefore, the best fitting model was the measurement model Ver2 (Figure 1) and it was used for further analysis.

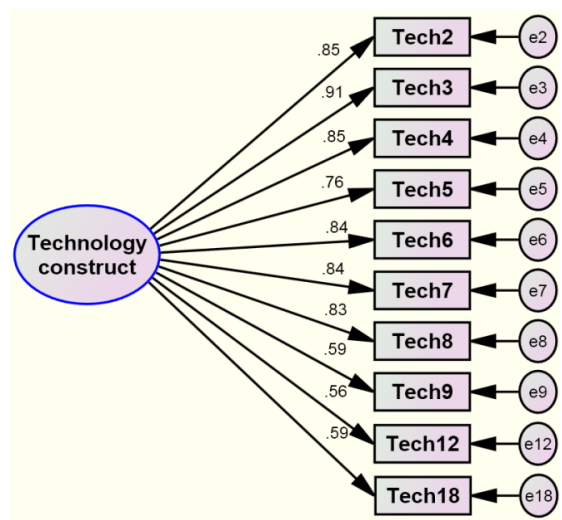

Figure 1 Measurement model Ver2

Table 4 Fitting indices (adopted from (Byrne, 2010))

| Fit Indices | Desired Range |
|---|---|
| χ2 /degrees of freedom (CMIN/DF) | ≤ 2.00 |
| IFI (Incremental Fit Index) | ≥ 0.90 |
| CFI (Comparative Fit Index) | Coefficient values range from zero to 1.00, with values close to .95 showing superior fit |
| RMSEA (Root Mean Square Error of Approximation) | values less than .05 show good fit, and values as high as .08 represent reasonable fit, from .08 to .10 show mediocre fit, and those greater than .10 show poor fit. |
| Root mean square residual (RMR) | ≤ 0.08 |
| Goodness-of-Fit Index (GFI) | ≥ 0.90 |
| Normed Fit Index (NFI) | Coefficient values range from zero to 1.00, with values close to .95 showing superior fit |
| Relative Fit Index (RFI) | Coefficient values range from zero to 1.00, with values close to .95 showing superior fit |
| Tucker-Lewis Index (TLI) | Values ranging from zero to 1.00, with values close to .95 (for large samples) being indicative of good fit. |

## 5 DISCUSSION ON VERIFIED MODEL

The final measurement was carried out based on measurement model ver2 by classifying the factors into two groups according to their relevant factor loading with a threshold value of 0.83. Referring to the Table 1, the proper name for each group can be Web-based communications and data sharing, respectively. From Figure 2, each factor loading with a value above 0.62 is significant. Overall, the final measurement model produced good fit indices (CMIN/DF = 2.889, RMR = 0.04, GFI = 0.929, RFI = 0.929, NFI = 0.949, TLI = 0.952, CFI = 0.966 IFI = 0.964, RMSEA = 0.089).

While fitting the information technology construct of the measurement model, the factors Tech14 (access shared files anytime, from any computer), Tech15 (web database), Tech16 (provide instant collaboration), Tech17 (software as a service (eliminating the need to install and run the application on the own computer) and Tech19 (can be integrated/compatible with the other tools and systems) were dropped. Modification indices (MI) based on regression weights showed that Tech17, Tech 18 and Tech19 were highly correlated, and therefore one representative (Tech18) from this group appeared to be adequate. Tech14 to Tech16 were strongly correlated with Tech12, and hence, the remaining factors represent the deleted ones.

The results of the final measurement model of information technology construct in virtual R&D team for new product development, showed the share of two main contrasts, which were strongly correlated to each other:

1. Web-based communications consists of online meetings on a required basis, web conferencing, seminars on the web, video conferencing, audio conferencing and online presentations.

2. Web-based data sharing consists of shared work spaces, shared documents (off-line), access service from any computer (in network) and virtual research centre for product development.

According to Lee-Kelley and Sankey (2008), these two constructs belong to the second and third generation technology. Well-equipped virtual R&D team members with the appropriate technology make the teams more effective. Therefore, managers of NPD should provide the facilities and infrastructures for the virtual R&D teams to achieve higher levels of team effectiveness.

Figure 2 Final measurement model

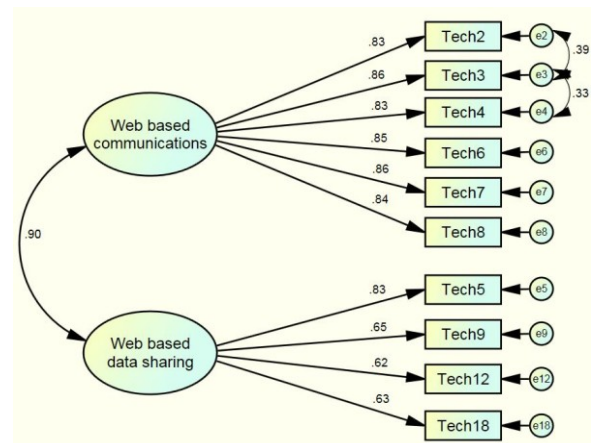

## 6 CONCLUSIONS

This research explores the 19 factors related to communication strategy using information technology in virtual team environment. However, the factors which mainly contribute to the information technology construct in virtual R&D teams' communication for new product development were unknown in the preceding literature. The findings of this study will contribute some knowledge in the literature and build a foundation for further understanding of the technology elements in virtual R&D teams for new product development. The measurement model shows ten factors that made the information technology constructs. These ten factors can be sorted by their factor loading, which reflects the factor's weight. Therefore, the software developer or the managers of NPD are able to provide a better platform for virtual teams by concentrating on the main factors. The second and third generation technologies (refer to definition of Lee-Kelley and Sankey (2008)) are now more suitable for developing new products through virtual R&D teams.

Future research is needed to examine the effects of each factor to perform the virtual R&D teams whereas the other constructs of virtual teams such as process and people are taken into account. A new SEM is needed to demonstrative the relationships between factors-construct and construct-construct, which are not yet investigated.

## 7 REFERENCES

ALE EBRAHIM, N., ABDUL RASHID, S. H., AHMED, S. & TAHA, Z. 2011. The Effectiveness of Virtual R&D Teams in SMEs: Experiences of Malaysian SMEs. *Industrial Engineering & Management Systems,* 10 (ARTICLE IN PRESS).

ALE EBRAHIM, N., AHMED, S., ABDUL RASHID, S. H. & TAHA, Z. Year. The Effectiveness of Virtual R&D Teams in SMEs: Experiences of Malaysian SMEs. *In:* The 11th Asia Pacific Industrial Engineering and Management Systems Conference 2010 (APIEMS 2010) December 7-10 2010 Melaka, Malaysia. Kuala Lumpur, Malaysia: University of Malaya Press, 1-6.

ALE EBRAHIM, N., AHMED, S. & TAHA, Z. 2009a. Modified Stage-Gate: A Conceptual Model of Virtual Product Development Process. *African Journal of Marketing Management,* 1, 211-219.

ALE EBRAHIM, N., AHMED, S. & TAHA, Z. 2009b. Virtual R & D teams in small and medium enterprises: A literature review. *Scientific Research and Essay,* 4**,** 1575–1590.

ALE EBRAHIM, N., AHMED, S. & TAHA, Z. Year. Virtual R&D Teams: Innovation and Technology Facilitator *In:* Engineering Education in 2025, 11-12 May 2009c School of Engineering and Technology, University of Tehran, Tehran, Iran. University of Tehran, 1-14.

ALE EBRAHIM, N., AHMED, S. & TAHA, Z. 2009d. Virtual Teams: a Literature Review. *Australian Journal of Basic and Applied Sciences,* 3**,** 2653-2669.

ANDERSON, A. H., MCEWAN, R., BAL, J. & CARLETTA, J. 2007. Virtual team meetings: An analysis of communication and context. *Computers in Human Behavior,* 23**,** 2558–2580.

ANDERSON, J. C. & GERBING, D. W. 1988. Structural equation modeling in practice: A review and recommended two-step approach. *Psychological Bulletin* 103**,** 411-423.

BASKERVILLE, R. & NANDHAKUMAR, J. 2007. Activating and Perpetuating Virtual Teams: Now That We're Mobile, Where Do We Go? *IEEE Transactions on Professional Communication,* 50**,** 17 - 34

BYRNE, B. M. 2010. *Structural equation modeling with AMOS: Basic concepts, applications, and programming,* New York, Taylor and Francis Group, LLC.

CHEN, M., LIOU, Y., WANG, C. W., FAN, Y. W. & CHI, Y. P. J. 2007. Team Spirit: Design, implementation, and evaluation of a Web-based group decision support system. *Decision Support Systems,* 43**,** 1186–1202.

CHUDOBA, K. M., WYNN, E., LU, M., WATSON-MANHEIM & BETH, M. 2005. How virtual are we? Measuring virtuality and understanding its impact in a global organization. *Information Systems Journal,* 15**,** 279-306.

COLEMAN, D. & LEVINE, S. 2008. *Collaboration 2.0: Technology and Best Practices for Successful Collaboration in a Web 2.0 World,* Silicon Valley, California, USA, Happy About®.

DEKKER, D. M., RUTTE, C. G. & VAN DEN BERG, P. T. 2008. Cultural differences in the perception of critical interaction behaviors in global virtual teams. *International Journal of Intercultural Relations,* 32**,** 441-452.

DIBRELL, C., DAVIS, P. S. & CRAIG, J. 2008. Fueling Innovation through Information Technology in SMEs. *Journal of Small Business Management,* 46**,** 203-

218.

EBRAHIM, N. A., AHMED, S. & TAHA, Z. 2010. Virtual R&D teams and SMEs growth: A comparative study between Iranian and Malaysian SMEs. *African Journal of Business Management,* 4**,** 2368-2379.

HUANG, C.-C. 2009. Knowledge sharing and group cohesiveness on performance: An empirical study of technology R&D teams in Taiwan. *Technovation,* 29**,** 786-797.

KOTELNIKOV, V. 2007. Small and Medium Enterprises and ICT. *In:* HAK-SU, K. (ed.) *Asia-Pacific Development Information Programme (UNDP-APDIP) e-Primers for the Information Economy, Society and Polity.* Bangkok: UNDP Regional Centre.

KRATZER, J., LEENDERS, R. & ENGELEN, J. V. 2005. Keeping Virtual R&D Teams Creative. *Research Technology Management,* 1**,** 13-16.

LAM, P.-K., CHIN, K.-S., YANG, J.-B. & LIANG, W. 2007. Self-assessment of conflict management in client-supplier collaborative new product development. *Industrial Management & Data Systems,* 107**,** 688 - 714.

LEE-KELLEY, L. & SANKEY, T. 2008. Global virtual teams for value creation and project success: A case study. *International Journal of Project Management* 26**,** 51–62.

LEENDERS, R. T. A. J., ENGELEN, J. M. L. V. & KRATZER, J. 2003. Virtuality, communication, and new product team creativity: a social network perspective. *Journal of Engineering and Technology Management,* 20**,** 69–92.

LIN, C., STANDING, C. & LIU, Y.-C. 2008. A model to develop effective virtual teams. *Decision Support Systems,* 45**,** 1031-1045.

LUREY, J. S. & RAISINGHANI, M. S. 2001. An empirical study of best practices in virtual teams *Information & Management,* 38**,** 523-544.

MCNAMARA, K., DENNIS, A. R. & CARTE, T. A. 2008. It's the Thought that Counts: The Mediating Effects of Information Processing in Virtual Team Decision Making. *Information Systems Management* 25**,** 20–32.

PAULEEN, D. J. & YOONG, P. 2001. Facilitating virtual team relationships via Internet and conventional communication channels. *Internet Research,* 11**,** 190 - 202.

PENA-MORA, F., HUSSEIN, K., VADHAVKAR, S. & BENJAMIN, K. 2000. CAIRO: a concurrent engineering meeting environment for virtual design teams. *Artifcial Intelligence in Engineering* 14**,** 203-219.

REDOLI, J., MOMPÓ, R., GARCÍA-DÍEZ, J. & LÓPEZ-CORONADO, M. 2008. A model for the assessment and development of Internet-based information and communication services in small and medium enterprises *Technovation,* 28**,** 424-435.

SHUMAROVA, E. V. 2009. *Authority-based and Bottom-up Diffusion of Collaboration Information Technologies: Constraints and Enablements.* MBA(Mgmt Information Systems), Universität Koblenz-Landau.

SILLS, S. J. & SONG, C. 2002. Innovations in Survey Research: An Application of Web-Based Surveys. *Social Science Computer Review,* 20**,** 22-30.

STOCK, G. N. & TATIKONDA, M. V. 2004. External technology integration in product and process development. *International Journal of Operations & Production Management,* 24**,** 642-665.

THISSEN, M. R., JEAN, M. P., MADHAVI, C. B. & TOYIA, L. A. 2007. Communication tools for distributed software development teams. *Proceedings of the 2007 ACM SIGMIS CPR conference on Computer personnel research: The global information technology workforce.* St. Louis, Missouri, USA: ACM.

TOWNSEND, A. M., DEMARIE, S. M. & HENDRICKSON, A. R. 1998. Virtual Teams: Technology and the Workplace of the Future. *The Academy of Management Executive,* 12**,** 17-29.

WALVOORD, A. A. G., REDDEN, E. R., ELLIOTT, L. R. & COOVERT, M. D. 2008. Empowering followers in virtual teams: Guiding principles from theory and practice. *Computers in Human Behavior,* 24**,** 1884-1906.

ZEMLIANSKY, P. & AMANT, K. S. 2008. *Handbook of Research on Virtual Workplaces and the New Nature of Business Practices,* New York, IGI Global; illustrated edition edition (April 7, 2008).